# Neutralizing Optical Defects in GeSn

Nirosh M. Eldose[1,a,‡], Dinesh Baral[1,2,a,‡], Diandian Zhang[1], Fernando Maia de Oliveira[1], Hryhorii Stanchu[1], Mohammad Zamani-Alavijeh[1], Yuriy I. Mazur[1], Wei Du[1,2], Shui-Qing Yu[1,2], and Gregory J. Salamo[1]

[1]*Institute for Nanoscience and Engineering, University of Arkansas, Fayetteville, Arkansas, USA*
[2]*Department of Electrical Engineering and Computer Science, University of Arkansas, Fayetteville, Arkansas, USA*

[a]These authors contribute equally.
[‡]Corresponding Authors: NME: nmeckama@uark.edu; DB: dbaral@uark.edu

**Abstract**

Reports of photoluminescence from GeSn grown on Ge substrates by molecular beam epitaxy have been limited. We find that one limiting factor to observing photoluminescence is due to localized defect states marked by photoluminescence at 2400 nm and originating from the Ge substrate and buffer layer. In this study, we report on an optical study utilizing doped Ge(001) substrates to effectively suppress defect-related photoluminescence in GeSn layers by filling localized defect trap states. For this experiment, a GeSn layer with Sn content up to 10.5% was grown on a doped Ge(001) substrate. Analysis of the physics of the photoluminescence spectrum collected from the GeSn thin film shows an emission at the expected wavelength of 2300 nm for 10.5% Sn content and the absence of the typically observed defect related signal at 2400 nm. This understanding is further confirmed using short pulse optical excitation of the GeSn grown on undoped Ge substrates.

**Keywords**: Germanium tin, Photoluminescence, Molecular beam epitaxy, Chemical vapor deposition.

## 1. Introduction

The development of group-IV semiconductor materials presents an exciting opportunity to advance silicon- and germanium-based photonics[1,2]. For example, germanium tin (GeSn) and silicon germanium tin (SiGeSn) alloys have the ability to transition Ge from an indirect to a direct bandgap for tin (Sn) concentrations between 6% and 10% depending on strain[3–13]. This unique property makes GeSn and SiGeSn promising candidates for mid-infrared (MIR) light emission, photodetection, and silicon-integrated optical applications[7,9]. However, while these are exciting possibilities with potential for significant impact, due to defects generated during the growth, achieving high-quality and stable GeSn layers with sufficiently high Sn incorporation remains a significant challenge[14–16].

Growth of GeSn has been explored using both chemical vapor deposition (CVD) and molecular beam epitaxy (MBE)[17,18]. For example, the growth of GeSn and SiGeSn on Ge/Si substrate by CVD has been successful resulting in both laser[19,20] and detector[21][22–25] demonstrations. However, even in this case, there is still need for improvement to reduce defect formation and suppress Sn segregation[26]. Although less studied and with less success when compared to CVD, MBE has presented an opportunity for a complementary approach with greater control over the growth environment, atomic-layer precision, flux of the individual material components, and growth temperatures[27]. For example, MBE is potentially a good choice for the fabrication of device structures that require sharp interfaces and low defect densities, such as, cascade lasers[28]. As a result, there have been many investigations of the growth of GeSn on Ge by MBE. Ge has been selected over Ge/Si substrate likely to provide the opportunity for higher quality GeSn. However, this expectation has not been realized, instead the optical performance of MBE-grown GeSn on Ge has yet to match that of their CVD-grown counterparts[29]. This is evidenced by the lack of reports demonstrating photoluminescence (PL) emission from GeSn grown on Ge emphasizing the need for further investigation into the optical properties of MBE grown GeSn.

In this paper we report on identifying defects originating from the Ge substrate that penetrate a GeSn thin film top layer as at least one of the major factors limiting the PL emission from GeSn grown by MBE. By identifying defects in the substrate as a limiting factor we have also been able to overcome this limitation by neutralizing the capability of defect states in Ge to diminish PL emission from Ge and GeSn.

## 2. Experiment

To investigate the lack of PL from GeSn grown on Ge substrates we first investigated the Ge substrate. The optical spectrum of PL emission from an undoped Ge (001) substrate at 10 K excited with laser light at 532 nm revealed an expected PL signal due to the indirect bandgap transition at 1770 nm in Ge, but also an emission at 2400 nm (Fig. 1d) which we hypothesize as due to localized defects in Ge (Fig. 1a). The same defect PL emission at 2400 nm has been observed, but enhanced, in Ge buffer layer and GeSn thin film layer grown on Ge substrate indicating the need to understand the defect in the Ge substrate[30]. Assuming this emission was due to localized defects originating in the Ge substrate (Fig. 1a) we compared the PL emission spectrum from undoped Ge(001) substrate with both n- and p-type doped Ge substrates (Figs. 1d-f), with the intention to probe if doping would result in filled defect trap states at low temperature and remove their active presence at 2400 nm in the PL spectrum.

As anticipated, we observed that the PL spectrum from n-type doped Ge(001) substrates resulted in observation of a stronger indirect bandgap transition signal at 1770 nm but did not show the defect emission at 2400 nm. The PL defect emission is apparently absent as hypothesized due to filling trap states near the conduction band at 10 K, increasing recombination of excited carriers between conduction and valence bands and reducing recombination of excited carriers mediated by localized trap sites (Fig. 1c). Meanwhile, we also observed that the PL spectrum from p-type doped Ge(001) substrates showed both a higher indirect transition to the conduction band Ge and a weaker defect signal at 2400 nm than for undoped. Given our hypothesis for the role of defect states, this observation can be expected since p-type doping will increase the probability for the indirect transition due to the larger hole density in the valence band and consequently reduce the competitive probability of emission at localized trap sites. (Fig. 1b). To further examine our hypothesis we also investigated PL emission from MBE grown GeSn on undoped, n-type and p-type doped substrates with similar results including the total absence of PL emission from the defect sites at 2400 nm for n-type doped Ge and the observation of PL at 2300 nm from the conduction band to valence band transition expected for 10% Sn content strained to the substrate.

## 3. Model

Based on our observations of PL emission from doped and undoped Ge substrates, with and without an added Ge buffer and GeSn layer, we can understand the lack of reports of observation of PL from GeSn grown on Ge substrates by MBE. Fig. 1 illustrates the proposed mechanism by which the type of doping influences the optical emission characteristics of Ge. In the undoped regime (Fig. 1a), at low temperature, carriers excited high into the conduction band transfer down into the L-valley where they make an indirect transition to the valence band (~1770 nm) and to the lower energy defect trap states resulting in longer-wavelength, dual PL emission at ~2200 nm and ~2400 nm (Fig. 1d).

With p-type doped substrates (Fig. 1c), the valence band becomes occupied by holes which will favor the probability for an indirect bandgap transition from conduction band to the valence band, thus reducing transfer through defect levels. As a result, the relative intensity of defect assisted emission diminishes, and the indirect PL emission at 1770 nm becomes stronger.

For the n-type doped Ge substrate (Fig. 1b), defect levels have an even higher probability of being populated with carriers at 10 K, thus reducing the probability of transfer of carriers to the defect levels and therefore radiative or non-radiative recombination via defect states while increasing the indirect emission even more. This effectively neutralizes defect-assisted emission in PL, yielding dominant indirect band-edge emission (~1770 nm). The complete elimination of defect-derived signals in the PL spectrum provides strong evidence that the n-type doped Ge (001) substrate could effectively passivate by saturation of the defect levels. Based on these results, two GeSn samples were grown using MBE for study: (i) GeSn on an n-type doped Ge(001), and (ii) GeSn on an undoped Ge(001).

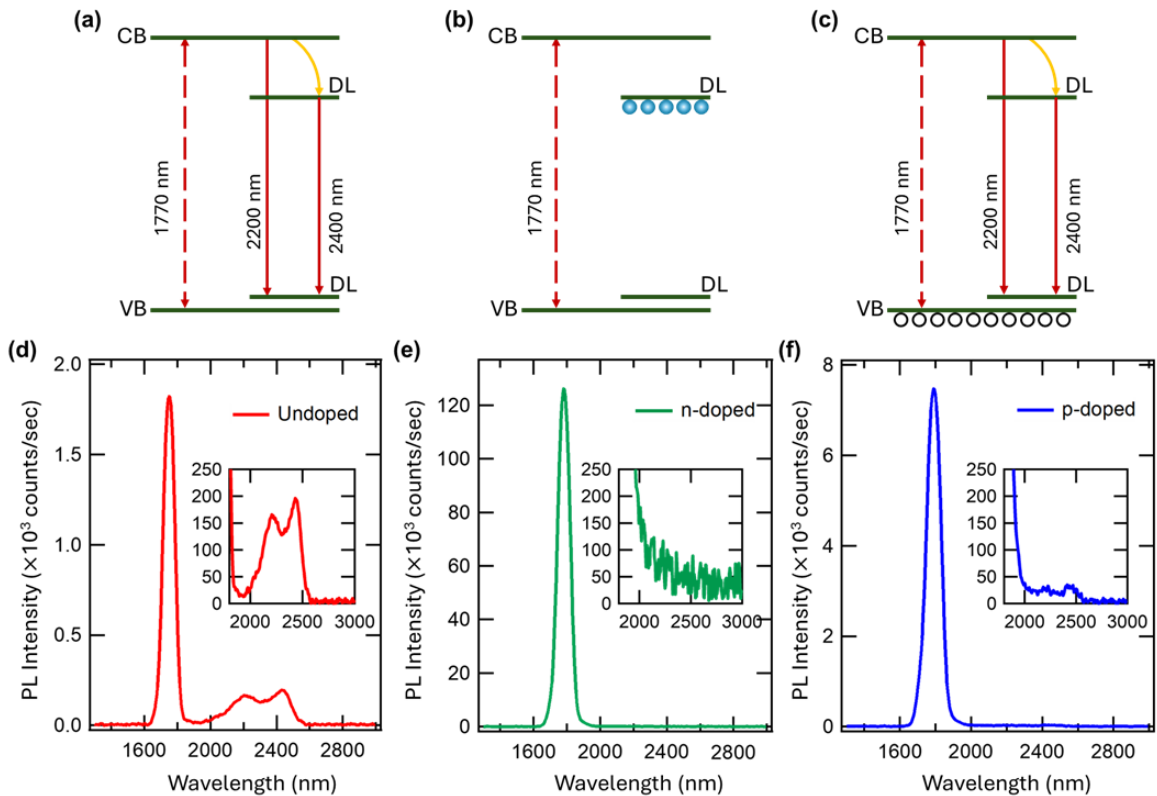

**Figure 1.** Schematic band diagrams illustrate carrier recombination pathways in Ge substrates with varying doping types. This model is also valid for GeSn. (a) Undoped material, where carriers recombine via direct band-edge and defect level (DL) transitions. (b) n-type doped material, with partial DL filling reducing defect-related emission. (c) p-type doped material, with fully occupied DLs yielding dominant direct band-edge emission. (d-f) PL emission from the undoped, n-type and p-type doped Ge(100) substrate at 10 K. Inset: show clear view of longer wavelength emission from the defect levels.

## 4. Growth

The growth of Ge buffers and GeSn films was performed using an MBE system equipped with Knudsen cells containing pyrolytic boron nitride (PBN) crucibles loaded with ultra-high-purity (7N) intrinsic Ge and (6N) metallic Sn sources. All the growths were conducted in an ultra-high vacuum (UHV) chamber with a base pressure of $10^{-11}$ mbar. Prior to growth, Ge wafers were chemically cleaned with diluted $HF:H_2O$ (1:20) mixtures for four minutes to remove the surface oxide followed by dip into $H_2O_2$ and $HCl:H_2O$ (1:4) for 30 seconds and one minute respectively for three cycles to obtain Cl-terminated surface[31]. The chemically cleaned Ge substrates were transferred quickly into the preparation chamber and preheated at 350 °C for 2 hours to desorb Cl termination. Next, it is heated at 750 °C for 1 hour to effectively desorb the $GeO_x$ passivation layer formed during the wet cleaning. A 200 nm Ge buffer layer was deposited at 450 °C to establish a clean and defect-free surface suitable for epitaxial growth of GeSn. Subsequently, GeSn layers were grown at a substrate temperature of 140 °C to enhance Sn incorporation.

## 5. Characterization

X-ray diffraction (XRD) measurements were performed using a Rigaku SmartLab Multipurpose X-ray Diffraction System equipped with a 3.0 kW Cu X-ray tube, four-bounce Ge(220) monochromator, 10.0 mm height limiting slit, and the HyPix-3000 pixel array multi-dimensional detector. The surface morphology was investigated by tapping mode atomic force microscopy (AFM) using a D3100 Nanoscope V scanning probe microscope (Bruker AXS, former Digital Instruments). The PL measurements were performed using a 220 mW 532 nm laser as excitation source and a Horiba spectrometer coupled with a PbS photodetector. A 1200 nm long-pass filter was used to block light from the laser excitation beam.

## 6. Results

Fig. 2a and 2b show the XRD 004 2θ-ω curve and reciprocal space maps (RSM), respectively, of GeSn layer grown on n-type doped Ge(001) substrate. On the 2θ/ω scan, the GeSn peak exhibits pendellösung fringes indicating good crystal quality. According to the distance between the oscillations, the GeSn layer thickness is 120 nm for both samples grown on n-type doped as well as undoped Ge substrates, which is close to the critical thickness for strain relaxation for GeSn[32]. On the RSMs, the $Ge_{1-x}Sn_x$ peak is vertically aligned with Ge substrate peak, indicating that the $Ge_{1-x}Sn_x$ layer grows pseudomorphically to Ge without significant strain relaxation. This is evidenced by $Q_{||}$ coordinate of the $(\bar{2}\bar{2}4)$ reciprocal lattice point which reflects the in-plane lattice parameter ($a$) by $Q_{||} = 2\pi\sqrt{8}/a$. The GeSn peak position along the $c$-axis is due to both the composition and the strain according to the following expressions,

$$\varepsilon_\perp = -2\frac{C_{12}}{C_{11}}\varepsilon_{||} \qquad (1)$$

$$c = a_0 - 2\frac{C_{12}}{C_{11}}(a - a_0) \qquad (2)$$

where $\varepsilon_\perp = (c - a_0)/a_0$, $\varepsilon_\parallel = (a - a_0)/a_0$, and $a_0$, $C_{12}$, and $C_{11}$ are the lattice and elastic constants of a $Ge_{1-x}Sn_x$ alloy that varies linearly with composition. The Sn content and in-plane strain ($\varepsilon_x$) in the GeSn layer were estimated from the reciprocal space position of the GeSn peak [33,34], and are 10.5% and -0.0152, respectively. Similar XRD-RSM data for a GeSn layer grown on an undoped Ge(001) substrate is shown in Fig. 2(c and d). This sample exhibits comparable high quality to the one grown on n-type doped Ge(001) substrate. Using the same calculation method described above, the Sn content was estimated to be 8.0%, with a strain value of -0.0116. This comparative sample was included with the specific purpose of providing a clear basis for comparison between the doped and undoped substrate, a motif of that will be discussed in detail later.

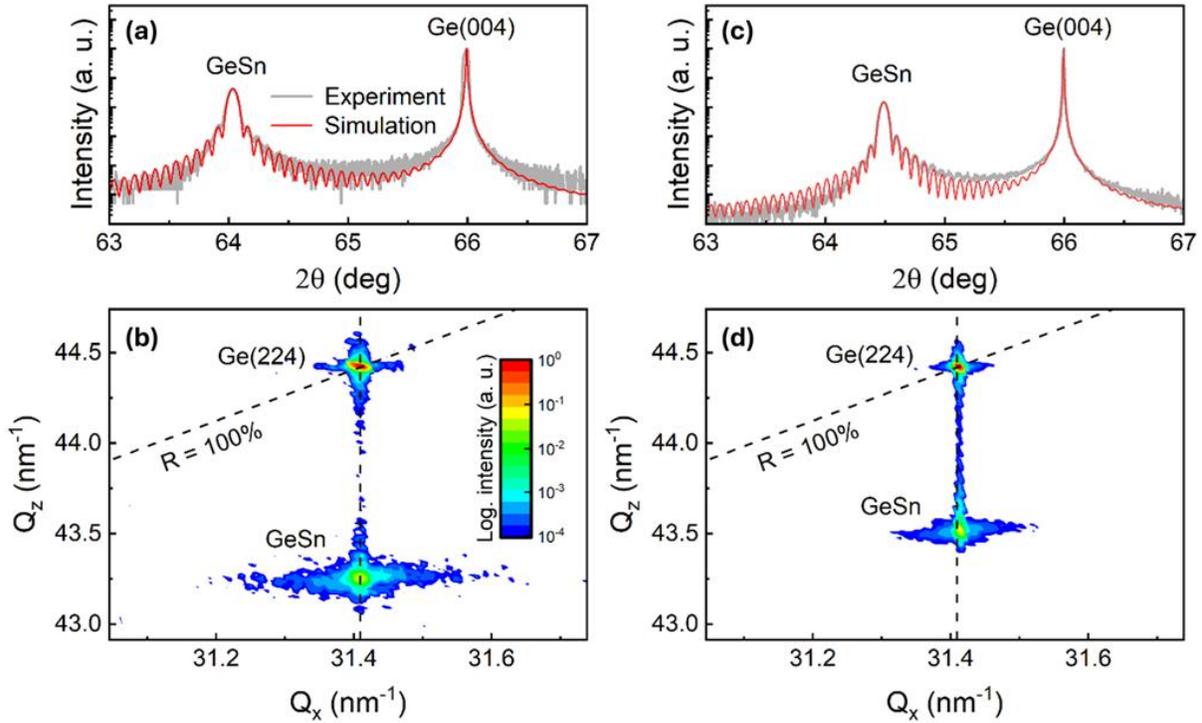

**Figure 2** (a) X-ray diffraction ω/2θ scans of symmetrical 004 reflection for GeSn layer with Sn content of 10.5% grown on n-type doped Ge(001) (b) shows its XRD-RSM collected near the asymmetrical ($\bar{2}\bar{2}4$) reflection of Ge. (c) X-ray diffraction ω/2θ scans of symmetrical 004 reflection for GeSn layer with Sn content of 8% grown on undoped Ge(001) (d) shows its XRD-RSM collected near the asymmetrical ($\bar{2}\bar{2}4$) reflection of Ge. The vertical dashed line in (b) and (d) corresponds to the fully strained $Ge_{1-x}Sn_x$ alloy.

Fig. 3a shows the AFM image (2 μm × 2 μm) of the Ge buffer grown on a n-type doped Ge substrate, which exhibits a relatively smooth surface. In comparison, Fig. 3b presents the AFM image of the GeSn layer with 10.5% Sn also grown on a Ge-buffer/Ge(001)(n-type) substrate. Similarly, Fig 3c presents the AFM image of GeSn layer with 8% Sn grown on the Ge-buffer/Ge(001)(undoped) substrate. Notably, the surface of both GeSn samples is free from significant Sn segregation, indicating a uniform Sn distribution achieved through pseudomorphic growth. X-ray photoelectron spectroscopy (XPS) of the surface indicated 10.5% GeSn in agreement with XRD. The root-mean-square (RMS) roughness of the GeSn layer is approximately 1.25 nm for the sample grown on n-type Ge, and 0.88 nm for the sample grown on undoped Ge.

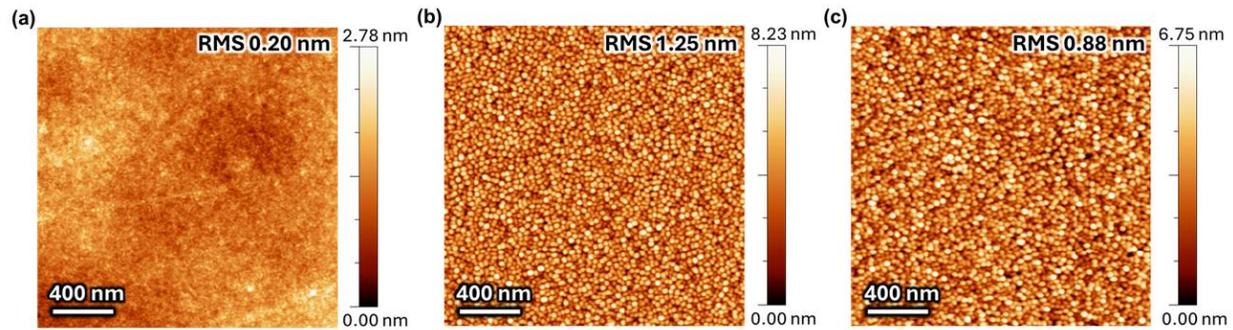

**Figure 3** shows AFM image (2 μm×2 μm) of (a) Ge buffer, (b and c) shows the $Ge_{1-x}Sn_x$ layer grown on n-type doped and undoped Ge(001) substrate respectively.

Fig. 4a presents the PL spectrum captured from a GeSn/Ge/Ge(001)(n-type) sample at 10 K. A single emission is seen at about 2300 nm. The PL spectrum of the Ge buffer/Ge(001) and bare n-type doped Ge substrate are shown in Figs. 4b-c, respectively, displaying a PL signal around 1775 nm, associated with the indirect bandgap of Ge at 10 K. This signal is not seen on the GeSn sample (Fig. 4a) since the penetration depth of the 532 nm probing laser limits the measured region to approximately the first 20 nm under the surface of the GeSn film, which is much shorter than the 120 nm thickness of the film. This approach allows us to characterize the GeSn film with minimum influence of the underlying buffer and substrate layers. Additionally, by choosing an n-type doped Ge (001) substrate we effectively neutralized the role of the 2400 nm defect-related states in Ge at low temperatures by adding carriers to the defect states. The high quality of the PL emission is also seen in Figs. 4d-j, which shows the PL signal persisting throughout the wide temperature range from 10 K to 300 K while maintaining about 70% of its initial integrated intensity when going from low to high temperatures.

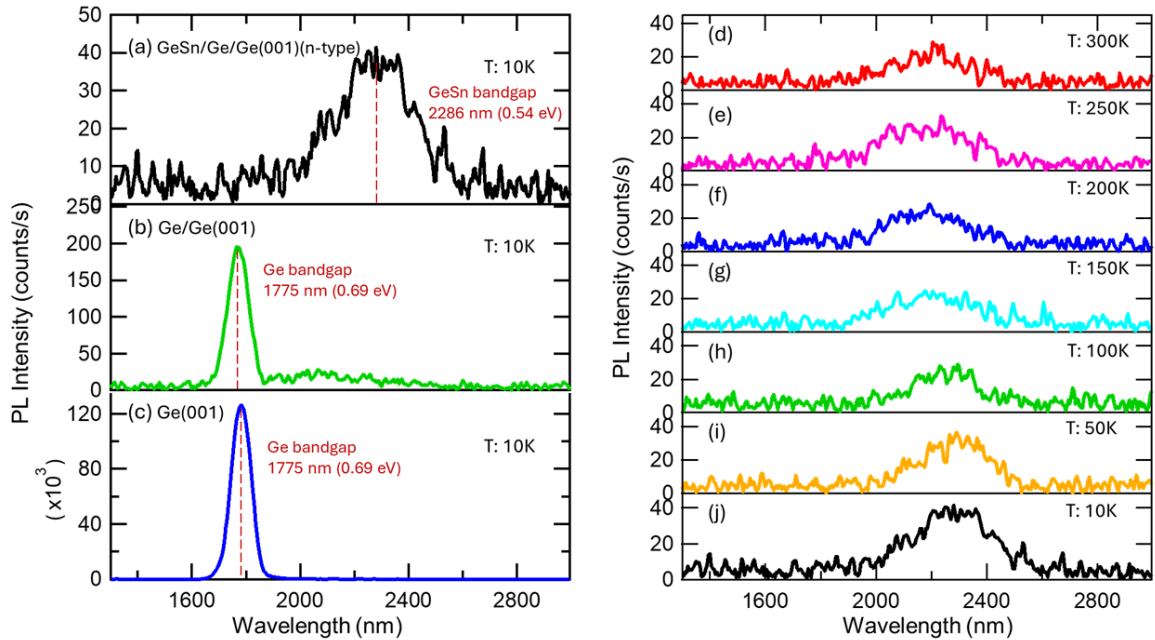

**Figure 4** (a) shows the PL emission from the GeSn layer at 10 K using 532 nm continuous laser (b) Ge buffer PL emission at 10K using 532 nm laser (c) n-type doped Ge (100) substrate reference PL emission (d-j) Temperature dependence of PL emission of the GeSn layer from 300 K to 10 K.

These results created an additional suggestion for us to investigate the use of a short optical laser pulse at low repetition rate (20 kHz) but high power (190 mW) to examine GeSn grown on an undoped Ge substrate for which we observed PL at 2400 nm but no PL due to the GeSn with about 8% Sn content when it was excited by a continuous laser. The idea was that the high optical intensity would saturate the defect state, and therefore we would observe PL from GeSn conduction band to valence band. Figure 5 shows the PL spectrum in this case, which displays the GeSn bandgap emission at about 2000 nm, further supporting the proposed model for a PL limiting defect state in GeSn grown on bulk Ge.

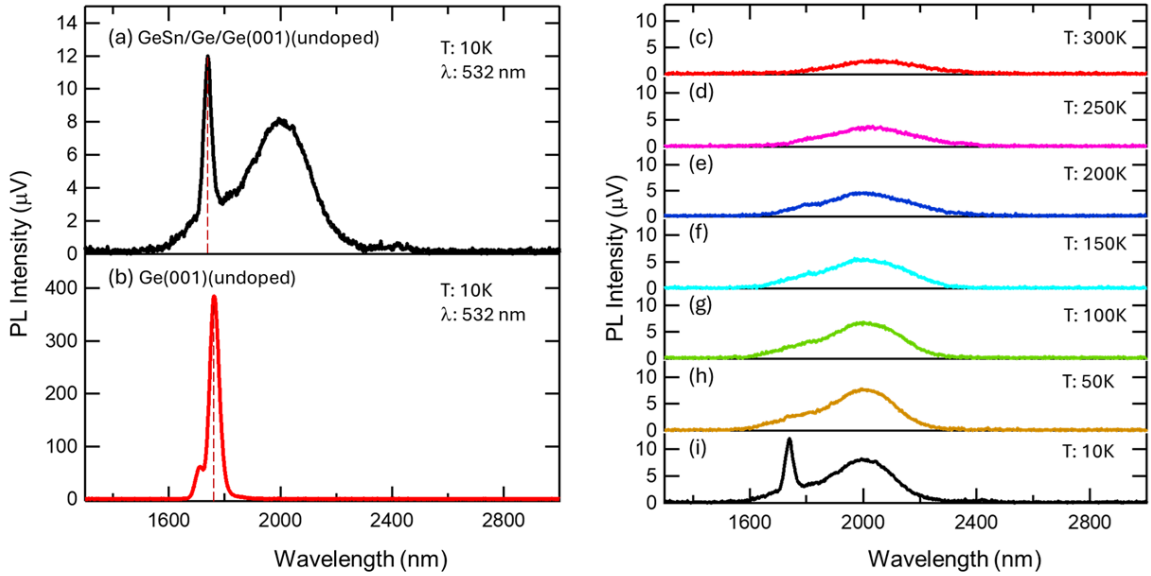

**Figure 5** (a) shows the PL emission from the GeSn layer at 10 K using 532 nm pulsed laser (b) undoped Ge (100) substrate reference PL emission (c-i) Temperature dependence of PL emission of the GeSn layer from 300 K to 10 K.

**Conclusion**

In this work, we have investigated the physics behind the optical emission from defects in Ge substrates and GeSn films with 8-10.5% Sn content. Importantly, low-temperature PL measurements revealed a defect-related emission feature near 2400 nm, which was identified as resulting from local defects originating in the Ge substrate. By utilizing an n-type Ge substrates, this defect emission was effectively neutralized, highlighting the critical role of substrate choice in achieving high-quality GeSn layers. Growth of GeSn on n-type Ge substrates was then used to demonstrate PL emission from GeSn with 10.5% Sn content and AFM analyses confirmed no evidence of significant Sn segregation. These findings underscore a pathway for optimizing GeSn growth and substrate engineering to potentially improve the optical quality of GeSn and SiGeSn structures grown in MBE or CVD.

**Acknowledgement**

This work was supported by the µ-ATOMS, an Energy Frontier Research Center funded by the U.S. Department of Energy, Office of Science, Basic Energy Sciences under award DE-SC0023412. This work was supported in part by the Office of Naval Research (Grant No. N00014-23-1-2872 and N00014-24-1-2651)

**Conflict of Interest**

There is no conflict of interest.

**CRediT authorship contribution statement**

**Nirosh M. Eldose:** Writing – original draft, Visualization, Methodology, Investigation, Validation. **Dinesh Baral:** Writing – original draft, Visualization, Methodology, Investigation, Validation. **Diandian Zhang:** Investigation. **Fernando Maia de Oliveira:** Data curation, Methodology, Investigation. **Hryhorii Stanchu:** Data Curation, Methodology, Investigation. **Mohammad Zamani-Alavijeh:** Investigation. **Yuriy I. Mazur:** Supervision. **Wei Du:** Supervision. **Shiu-Qing Yu:** Resources, Supervision, Project Administration, Funding Acquisition. **Gregory J. Salamo:** Writing – review & editing, Conceptualization, Resources, Supervision, Project administration, Funding Acquisition.

**References:**

[1] S. Wirths, D. Buca, and S. Mantl, "Si–Ge–Sn alloys: From growth to applications," Progress in Crystal Growth and Characterization of Materials **62**, 1–39 (2016).

[2] A. Giunto, and A. Fontcuberta i Morral, "Defects in Ge and GeSn and their impact on optoelectronic properties," Appl Phys Rev **11**, 041333 (2024).

[3] A. Mosleh, S.A. Ghetmiri, B.R. Conley, M. Hawkridge, M. Benamara, A. Nazzal, J. Tolle, S.Q. Yu, and H.A. Naseem, "Material characterization of Ge1-x Sn x alloys grown by a commercial CVD system for optoelectronic device applications," J Electron Mater **43**, 938–946 (2014).

[4] Y. Zhou, W. Dou, W. Du, S. Ojo, H. Tran, S.A. Ghetmiri, J. Liu, G. Sun, R. Soref, J. Margetis, J. Tolle, B. Li, Z. Chen, M. Mortazavi, and S.Q. Yu, "Optically Pumped GeSn Lasers Operating at 270 K with Broad Waveguide Structures on Si," ACS Photonics **6**, 1434 (2019).

[5] J. Margetis, S. Al-Kabi, W. Du, W. Dou, Y. Zhou, T. Pham, P. Grant, S. Ghetmiri, A. Mosleh, B. Li, J. Liu, G. Sun, R. Soref, J. Tolle, M. Mortazavi, and S.-Q. Yu, "Si-Based GeSn Lasers with Wavelength Coverage of 2−3 µm and Operating Temperatures up to 180 K," ACS Photonics **5**, 827 (2018).

[6] J. Aubin, J.M. Hartmann, A. Gassenq, J.L. Rouviere, E. Robin, V. Delaye, D. Cooper, N. Mollard, V. Reboud, and V. Calvo, "Growth and structural properties of step-graded, high Sn content GeSn layers on Ge," Semicond Sci Technol **32**, 094006 (2017).

[7] O. Moutanabbir, S. Assali, X. Gong, E.O. Reilly, J. Witzens, W. Du, A. Chelnokov, D. Buca, C.A. Broderick, D. Nam, B. Marzban, J. Witzens, W. Du, S.-Q. Yu, A. Chelnokov, D. Buca, and D. Nam, "Monolithic infrared silicon photonics : The rise of ( Si ) GeSn semiconductors," Appl Phys Lett **118**, 110502 (2021).

[8] B. Wang, E. Sakat, E. Herth, M. Gromovyi, A. Bjelajac, J. Chaste, G. Patriarche, P. Boucaud, F. Boeuf, N. Pauc, V. Calvo, J. Chrétien, M. Frauenrath, A. Chelnokov, V. Reboud, J.M. Hartmann, and M. El Kurdi, "GeSnOI mid-infrared laser technology," Light Sci Appl **10**, 232 (2021).

[9] V. Reboud, A. Gassenq, N. Pauc, J. Aubin, L. Milord, Q.M. Thai, M. Bertrand, K. Guilloy, D. Rouchon, J. Rothman, T. Zabel, F. Armand Pilon, H. Sigg, A. Chelnokov, J.M. Hartmann, and V. Calvo, "Optically pumped GeSn micro-disks with 16% Sn lasing at 3.1 µm up to 180 K," Appl Phys Lett **111**, 092101 (2017).

[10] L. Jiang, J.D. Gallagher, C.L. Senaratne, T. Aoki, J. Mathews, J. Kouvetakis, and J. Menéndez, "Compositional dependence of the direct and indirect band gaps in Ge$_{1-y}$Sn$_y$ alloys from room temperature photoluminescence: Implications for the indirect to direct gap crossover in intrinsic and n-type materials," Semicond Sci Technol **29**, 115028 (2014).

[11] S. V. Kondratenko, Y. V. Hyrka, Y.I. Mazur, A. V. Kuchuk, W. Dou, H. Tran, J. Margetis, J. Tolle, S.Q. Yu, and G.J. Salamo, "Photovoltage spectroscopy of direct and indirect bandgaps of strained Ge$_{1-x}$Sn$_x$ thin films on a Ge/Si(001) substrate," Acta Mater **171**, 40 (2019).

[12] C. Eckhardt, K. Hummer, and G. Kresse, "Indirect-to-direct gap transition in strained and unstrained Sn$_x$Ge$_{1-x}$ alloys," Phys Rev B **89**, 165201 (2014).

[13] D. Zhang, N.M. Eldose, D. Baral, H. Stanchu, S. Acharya, F.M. De Oliveira, M. Benamara, H. Zhao, Y. Zeng, W. Du, G.J. Salamo, and S. Yu, "Direct Bandgap Photoluminescence of GeSn grown on Si(100) substrate by Molecular Beam Epitaxy Growth," ArXiv, arXiv:2505.04096 (2025).

[14] W. Dou, M. Benamara, A. Mosleh, J. Margetis, P. Grant, Y. Zhou, S. Al-Kabi, W. Du, J. Tolle, B. Li, M. Mortazavi, and S.Q. Yu, "Investigation of GeSn Strain Relaxation and Spontaneous Composition Gradient for Low-Defect and High-Sn Alloy Growth," Sci Rep **8**, 5640 (2018).

[15] H. Pérez Ladrón De Guevara, A.G. Rodríguez, H. Navarro-Contreras, and M.A. Vidal, "Ge$_{1-x}$Sn$_x$ alloys pseudomorphically grown on Ge(001)," Appl Phys Lett **83**, 4942 (2003).

[16] N. M. Eldose, H. Stanchu, S. Das, I. Bikmukhametov, C. Li, S. Shetty, Y.I. Mazur, S.-Q. Yu, and G.J. Salamo, "Strain-Mediated Sn Incorporation and Segregation in Compositionally Graded Ge$_{1-x}$Sn$_x$ Epilayers Grown by MBE at Different Temperatures," Cryst Growth Des **23**, 7737 (2023).

[17] J. Zheng, Z. Liu, C. Xue, C. Li, Y. Zuo, B. Cheng, and Q. Wang, "Recent progress in GeSn growth and GeSn-based photonic devices," Journal of Semiconductors **39**, 061006 (2018).

[18] Y. Miao, G. Wang, Z. Kong, B. Xu, X. Zhao, X. Luo, H. Lin, Y. Dong, B. Lu, L. Dong, J. Zhou, J. Liu, and H.H. Radamson, "Review of Si-based GeSn CVD growth and optoelectronic applications," Nanomaterials **11**, 2556 (2021).

[19] Y. Zhou, Y. Miao, S. Ojo, H. Tran, G. Abernathy, J.M. Grant, S. Amoah, G. Salamo, W. Du, J. Liu, J. Margetis, J. Tolle, Y. Zhang, G. Sun, R.A. Soref, B. Li, and S.-Q. Yu, "Electrically Injected GeSn Laser on Si Operating up to 110K," Optica **7**, 924 (2021).

[20] G. Abernathy, S. Ojo, A. Said, J.M. Grant, Y. Zhou, H. Stanchu, W. Du, B. Li, and S.Q. Yu, "Study of all-group-IV SiGeSn mid-IR lasers with dual wavelength emission," Sci Rep **13**, 18515 (2023).

[21] M.-H. Chou, R. Bansal, Y.-T. Jheng, G. Sun, W. Du, S.-Q. Yu, and G.-E. Chang, "High-Detectivity GeSn Mid-Infrared Photodetectors for Sensitive Infrared Spectroscopy," Adv Photonics Res **6**, 2400155 (2025).

[22] X. Li, L. Peng, Z. Liu, Z. Zhou, J. Zheng, C. Xue, Y. Zuo, B. Chen, and B. Cheng, "30 GHz GeSn photodetector on SOI substrate for 2 μm wavelength application," Photonics Res **9**, 494 (2021).

[23] T. Pham, W. Du, H. Tran, J. Margetis, J. Tolle, G. Sun, R.A. Soref, H.A. Naseem, B. Li, and S.-Q. Yu, "Systematic study of Si-based GeSn photodiodes with 26 µm detector cutoff for short-wave infrared detection," Opt Express **24**, 4519 (2016).

[24] H.H. Tseng, H. Li, V. Mashanov, Y.J. Yang, H.H. Cheng, G.E. Chang, R.A. Soref, and G. Sun, "GeSn-based p-i-n photodiodes with strained active layer on a Si wafer," Appl Phys Lett **103**, 231907 (2013).

[25] I. Dascalescu, C. Palade, A. Slav, I. Stavarache, O. Cojocaru, V.S. Teodorescu, V.A. Maraloiu, A.M. Lepadatu, M.L. Ciurea, and T. Stoica, "Enhancing SiGeSn nanocrystals SWIR photosensing by high passivation in nanocrystalline $HfO_2$ matrix," Sci Rep **14**, 3532 (2024).

[26] H. Stanchu, A. Said, O. Olorunsola, S. Acharya, S. Amoah, M. Zamani-Alavijeh, F.M. de Oliveira, S.K. Chhetri, J. Hu, Y.I. Mazur, S.-Q. Yu, and G. Salamo, "Role of dislocations on Sn diffusion during low temperature annealing of GeSn layers," Journal of Vacuum Science & Technology B **41**, 052208 (2023).

[27] A.B. Talochkin, and V.I. Mashanov, "Formation of GeSn alloy on Si(100) by low-temperature molecular beam epitaxy," Appl Phys Lett **105**, 263101 (2014).

[28] H. Nguyen-Van, A.N. Baranov, Z. Loghmari, L. Cerutti, J.B. Rodriguez, J. Tournet, G. Narcy, G. Boissier, G. Patriarche, M. Bahriz, E. Tournié, and R. Teissier, "Quantum cascade lasers grown on silicon," Sci Rep **8**, 7206 (2018).

[29] L. Zhang, Y. Song, N. von den Driesch, Z. Zhang, D. Buca, D. Grützmacher, and S. Wang, "Structural property study for GeSn thin films," Materials **13**, 3645 (2020).

[30] F. Maia de Oliveira, N. M. Eldose, D. Baral, H. Stanchu, S. Kryvyi, D.D. Zhang, M. Zamani-Alavijeh, M. Benamara, Y.I. Mazur, W. Du, S.Q. Yu, and G.J. Salamo, "Dislocation-Related Photoluminescence Emission in MBE-Grown GeSn," Cryst Growth Des **25**, 6023 (2025).

[31] N.M. Eldose, F.M. De Oliveira, S. Kryvyi, H. Stanchu, D. Baral, D. Zhang, S. Liu, Y. Wang, J. Liu, M. Benamara, M. Zamani-Alavijeh, I. Bikmukhametov, Y.I. Mazur, W. Du, S.-Q. Yu, and G.J. Salamo1, "MBE growth and characterization of strained GeSn / Ge multiple quantum well," J. Phys. D: Appl. Phys **58**, 225111 (2025).

[32] W. Wang, Q. Zhou, Y. Dong, E.S. Tok, and Y.C. Yeo, "Critical thickness for strain relaxation of $Ge_{1-x}Sn_x$ ($x \leq 0.17$) grown by molecular beam epitaxy on Ge(001)," Appl Phys Lett **106**, 232106 (2015).

[33] H. V. Stanchu, A. V. Kuchuk, Y.I. Mazur, K. Pandey, F.M. De Oliveira, M. Benamara, M.D. Teodoro, S.Q. Yu, and G.J. Salamo, "Quantitative Correlation Study of Dislocation Generation, Strain Relief, and Sn Outdiffusion in Thermally Annealed GeSn Epilayers," Cryst Growth Des **21**, 1666 (2021).

[34] J. Aubin, J.M. Hartmann, A. Gassenq, L. Milord, N. Pauc, V. Reboud, and V. Calvo, "Impact of thickness on the structural properties of high tin content GeSn layers," J Cryst Growth **473**, 20 (2017).